\documentclass[prl,twocolumn,aps,superscriptaddress,showpacs]{revtex4}
\usepackage{graphics}
\usepackage{color}
\usepackage{graphicx}
\usepackage{bm}
\newcommand{\UQ}{School of Physical Sciences, University of Queensland, Brisbane, QLD 4072, Australia}
\newcommand{\ANU}{Department of Physics, Australian National University, Canberra, Australia}
\newcommand{\ARC}{Australian Research Council Centre of Excellence for Quantum-Atom Optics}
\newcommand{\EQ}[1]{\begin{eqnarray}#1\end{eqnarray}}

\newcommand{\e}{\mbox{e}}
\newcommand{\notes}[1]{}
\def\NOTES{
\renewcommand{\notes}[1]{\\{\color{red}\emph{##1}}\\}}
\NOTES
\begin{document}
\title{Measuring the quantum statistics of an atom laser beam}
\author{A.~S. Bradley}
\affiliation{\ARC} 
\affiliation{\UQ}
\author{M.~K. Olsen}
\affiliation{\ARC} 
\affiliation{\UQ}
\author{S.~A. Haine}
\affiliation{\ARC} 
\affiliation{\ANU}
\author{J.~J. Hope}
\affiliation{\ARC} 
\affiliation{\ANU}

\begin{abstract}
We propose and analyse a scheme for measuring the quadrature statistics of an atom laser beam using extant optical homodyning and Raman atom laser techniques. 
Reversal of the normal Raman atom laser outcoupling scheme is used to map the quantum statistics of an {\em incoupled} beam to an optical probe beam. A multimode model of the spatial propagation dynamics shows that the Raman incoupler gives a clear signal of de~Broglie wave quadrature squeezing for both pulsed and continuous inputs. Finally, we show that experimental realisations of the scheme may be tested with existing methods via measurements of Glauber's intensity correlation function.
\end{abstract}
\maketitle
Quantum-Atom Optics~\cite{Lenz1993,Rolston2002,Knight2005}, the study of quantum properties of matter waves, is a rapidly developing subfield of ultra-cold atomic physics.
Recent experimental progress includes measurements of intensity correlations of noncondensed $^{20}$Ne~\cite{Yasuda1996}, Hanbury Brown-Twiss correlations~\cite{Schellekens2005,Chuu2005}, fermion pairing correlations~\cite{Greiner2005}, spatial correlations of density fluctuations~\cite{Folling2005}, and sub-Poissonian number fluctuations~\cite{Ottl2005}. 
Despite the advances in cold atom detection techniques which have made such measurements possible, information available from intensity correlation experiments is restricted to correlation functions of the Glauber type~\cite{Glauber1963}. As is well known in quantum optics, probing quantum states generated by nonlinear interactions requires controllable phase-sensitive detection techniques. 
Optical quadrature variances, analogous to the quantum uncertainties of momentum and position of a particle, are measured via homodyne detection~\cite{Yuen1978,Collett1987}. This technique has been used to demonstrate optical squeezing~\cite{Slusher1985,Wu1986}, the Einstein-Podolsky-Rosen paradox for photons~\cite{Ou1992}, and continuous variable teleportation~\cite{Furusawa1998}, and is central to quantum information science~\cite{Braunstein2005}. Although interference~\cite{Andrews1997}, intensity correlation and tomographic measurements~\cite{Moore2006} have been performed with bosonic matter waves, a practical scheme to realise matter wave homodyne detection has not yet been demonstrated.

Proposed methods for producing matter waves in highly non-classical
states include utilising the nonlinear atomic interactions to create
correlated pairs of atoms via either molecular down conversion \cite{Kheruntsyan2005}, spin exchange collisions \cite{Duan2000b,Pu2000}, or by transferring the quantum state of a non-classical electromagnetic field to a propagating atomic field
\cite{Jing2000,Haine2005a,Fleischhauer2002}. In some of these schemes it has
been demonstrated that continuous variable entanglement can be
generated between spatially separated atomic beams \cite{Kheruntsyan2005, Haine2005a} or between an atomic beam and an
optical beam \cite{Haine2006}  which can be used to perform tests of quantum non-locality
with massive particles. Quadrature measurements on free
atomic fields will be necessary to observe these effects and although
schemes for atomic homodyne measurements have been proposed these are
confined to trapped BEC~\cite{Corney1998,Search2001,DaCunha2006}. 

%Theoretical proposals to produce matter waves in highly nonclassical states include photodissociation~\cite{Kostrun2000}, entangled atom laser %beams~\cite{Haine2005}, atom-light entanglement~\cite{Haine2006}, and a matter wave demonstration of the EPR paradox~\cite{Kheruntsyan2005}. 
%There have also been theoretical and experimental investigations of schemes to produce atomic samples exhibiting continuous variable entanglement, %particularly using four-wave mixing in a periodic potential~\cite{Hilligsoe2005,Campbell2006,Olsen2006a}. Quadrature measurements on free atomic fields will %be necessary to quantify all these effects and although schemes for atomic homodyne measurement have been proposed these are confined to trapped %BEC~\cite{Corney1998,Search2001,DaCunha2006}.
%===============================================
\begin{figure}\begin{centering}\includegraphics[width=1\columnwidth]{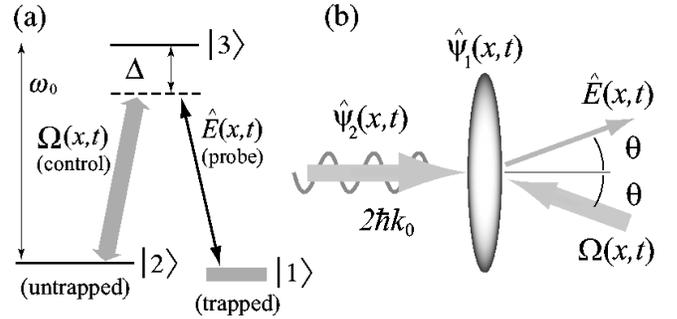}\par\end{centering}
\vspace*{-10pt}
\caption{\label{fig:vlife_schematic}\textbf{Schematic of a Raman atom laser incoupler}. (a) $\Lambda$-configuration of three level atoms. An untrapped beam of $|2\rangle$ atoms is coupled to a trapped state $|1\rangle$ via a Raman transition. The two optical fields are a weak probe beam (annihilation operator $\hat{E}(x,t)$), and a control beam ($\Omega(x,t)$), modelled by a classical field. The transition is detuned from the intermediate state by $\Delta$. Wide lines represent highly occupied states. (b) Spatial configuration of the Raman atom laser system. Beam atoms (field operator $\hat{\psi}_2(x,t)$) reach the condensate ($\hat{\psi}_{1}(x,t)$) with momentum $2\hbar k_0$, which is transferred during incoupling by absorption and emission of a light quanta with momenta $\mp\hbar k_0$ along the propagation axis.}
\vspace*{-10pt}
\end{figure}
%===============================================

In this Letter we propose a scheme for dynamically transferring quantum information from a propagating atom laser beam to an optical beam, allowing indirect measurement of de Broglie wave quadrature variances via optical homodyning. By analogy with optical quadratures, $\hat{X}(\theta)=\hat{a}\e^{-i\theta}+\hat{a}^{\dag}\e^{i\theta}$, we may use atomic field operators to define atomic quadratures~\cite{Kheruntsyan2005}, and the scheme which we analyse here is designed to measure the quadratures of a propagating atomic beam. It involves a reversal of the successful Raman atom laser output coupling scheme~\cite{Moy1997,Hagley1999}, a variant of which has previously been proposed as a mechanism for transferring states of a nonclassical optical field to the outgoing atomic beam~\cite{Haine2005a,Haine2005,Haine2006}. As we will show, a two-photon Raman transition allows the atom laser beam to be incoupled~\cite{Paranjape2003} to a large trapped condensate, with highly efficient transfer of the atomic statistics to an outgoing optical field. 

The scheme (Fig.~\ref{fig:vlife_schematic}) consists of a trapped condensate and an incoming atom laser beam of the same atomic species. The internal state Raman energy level configuration allows for stimulated transitions between the trapped and untrapped fields. These transitions are stimulated by two optical fields, one of which is intense (control) and denoted by its Rabi frequency $\Omega(x,t)$, while the other is much weaker (probe) and denoted by the field operator $\hat{E}(x,t)$. We perform our analysis using a one-dimensional model, described by the Hamiltonian ${\cal H}={\cal H}_{\rm atom}+{\cal H}_{\rm int}+{\cal H}_{\rm light}$, with
\EQ{
{\cal H}_{\rm atom}&=&\sum_{j=1}^3\int dx\;\hat{\psi}_j^\dag(x)H_j\hat{\psi}_j(x),\\
{\cal H}_{\rm int}&=&\hbar\int dx\;\left(\hat{\psi}_2(x)\hat{\psi}_3^\dag(x)\Omega(x,t)+h.c.\right)\nonumber\\
&&+\hbar g_{13}\int dx\;\left(\hat{E}(x)\hat{\psi}_1(x)\hat{\psi}_3^\dag(x)+h.c.\right),\\
{\cal H}_{\rm light}&=&\int dx\;\hat{E}^\dag(x)pc\hat{E}(x),
}
where $H_1=-\frac{\hbar^2\partial_x^2}{2m}+V_{1}(x)$, $H_2=-\frac{\hbar^2\partial_x^2}{2m}+V_{2}(x)$, $H_3=-\frac{\hbar^2\partial_x^2}{2m}+\hbar\omega_0+V_{3}(x)$, $m$ is the atomic mass, and the $V_{j}$ represent both linear (trapping for $\hat{\psi}_{1}$) and nonlinear (scattering) potentials. The optical control field is $\Omega(x,t)=\Omega_{23}e^{i(k_0x-(\omega_0-\Delta)t)}$ where $\Omega_{23}$ is the Rabi frequency for the $|2\rangle \to |3\rangle$ transition. $\hat{\psi}_1(x)$, $\hat{\psi}_2(x)$, $\hat{\psi}_3(x)$ and $\hat{E}(x)$ are the annihilation operators for the condensate mode (internal state $|1\rangle$), signal beam ($|2\rangle$), excited state atoms ($|3\rangle$), and probe beam photons respectively, satisfying the usual bosonic commutation relations, $[\hat{\psi}_i(x),\hat{\psi}_j^\dag(x^\prime)]=\delta_{ij}\delta(x-x^\prime)$ and $[\hat{E}(x),\hat{E}^\dag(x^\prime)]=\delta(x-x^\prime)$.
The coupling coefficient is $g_{13}=(d_{13}/\hbar)\sqrt{\hbar\omega_k/2\epsilon_0}$, where $d_{13}$ is the electric dipole moment for the $|1\rangle \to |2\rangle$ transition. We neglect interatomic interactions on the basis that the atomic beam is dilute and the process will take place over a time short enough that any phase diffusion effects will be minimal. 
We now introduce the rotating frame fields $\tilde{\psi}_3(x)=\hat{\psi}_3(x)e^{i(\omega_0-\Delta)t}$ and $\tilde{E}(x)=\hat{E}(x)e^{i(\omega_0-\Delta)t}$ and adiabatically eliminate the weakly occupied intermediate state~\cite{Haine2005,Haine2006}
$\tilde{\psi}_3(x)\to-\frac{\Omega_{23}}{\Delta}e^{ik_0x}\hat{\psi}_2(x)-\frac{g_{13}}{\Delta}\tilde{E}(x)\hat{\psi}_1(x)$.
We approximate the highly occupied condensate as a coherent state, $\hat{\psi}_{1}(x,t)=\phi(x,t)\equiv\langle\hat{\psi}_{1}(x,t)\rangle$, while allowing the occupation and the spatial shape to change. To simplify notation we set $\hat{\psi}_2\equiv\hat{\psi}$ to arrive at the equations of motion
\EQ{
\label{psieom}i\dot{\hat{\psi}}(x)&=&H_a\hat{\psi}(x)-\Omega_C(x)e^{-ik_0x}\tilde{E}(x)\\
\label{Eeom}i\dot{\tilde{E}}(x)&=&H_b\tilde{E}(x)-\Omega_C^*(x)e^{ik_0x}\hat{\psi}(x).\\
\label{phieom}i\dot{\phi}(x)&=&H_{\phi}\phi(x)-\frac{g_{13}\Omega_{23}}{\Delta}e^{ik_0x}\langle \hat{E}^\dag(x)\hat{\psi}(x)\rangle
}
%===============================================
\begin{figure}\begin{centering}\includegraphics[width=1\columnwidth]{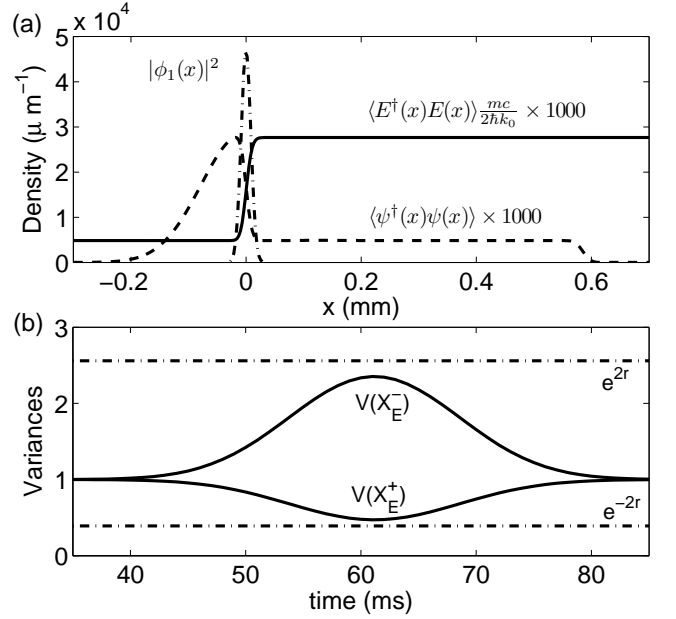}\par\end{centering}

\vspace*{-10pt}

\caption{\label{fig:packetTimeSeries}\textbf{Incoupling an atom laser pulse}. (a) A squeezed atomic pulse (4dB in the $X^+$-quadrature, dashed line) of $n_0=5\times 10^3$ atoms initially centered at $x=-600\mu {\rm m}$, with momentum wavevector $2k_0$, is coupled into the condensate (chain line). The probe field (solid line) has peak intensity occurring at $t=54$ms shown here, which is when the maximum of the pulse is centred on the condensate. The atom pulse and optical probe are magnified by factors of $1000$ and $1000\times mc/2\hbar k_0$ to plot them on the condensate scale. (b) Time development of the probe quadratures. A value of less than one demonstrates quadrature squeezing and the chain lines give the atomic variances.}

\vspace*{-10pt}

\end{figure}
%================================================
with $H_a=-\hbar\partial_x^2/2m-|\Omega_{23}|^2/\Delta$, $H_b=-ic\partial_x-|\phi(x)|^2(g_{13})^2/\Delta+\Delta-\omega_0$,
$H_{\phi}=-\hbar\partial_x^2/2m+V_1(x)/\hbar-\langle\hat{E}^\dag(x)\hat{E}(x)\rangle(g_{13})^2/\Delta$, and 
$\Omega_C(x)=\phi(x)\Omega_{23}^*g_{13}/\Delta$.
As shown in Ref.~\cite{Haine2005}, equations of this type can be efficiently solved to give all relevant observables.
\par 
We now define mode matched quadratures~\cite{Kheruntsyan2005} which characterise the probe light and the atomic signal. We define a mode of the atomic ($\nu=\psi$) or optical ($\nu=E$) field $L_\nu(x,t)$, and the operators
$
\hat{a}_\nu=\int_{x_1^\nu}^{x_2^\nu} dx\;L_\nu^*(x,t)\hat{\nu}(x,t)
$,
with the normalisation $\int_{x_1^\nu}^{x_2^\nu} dx\;L_\nu^*(x,t)L_\nu(x,t)=1$, satisfying $[\hat{a}_\nu,\hat{a}_{\nu^\prime}^\dag]=\delta_{\nu \nu^\prime}$. 
The mode matched quadratures
$
\hat{X}_\nu^+=\hat{a}_\nu+\hat{a}_\nu^\dag$, $\hat{X}_\nu^-=i(\hat{a}_\nu^\dag-\hat{a}_\nu)$
have commutator $[\hat{X}_\nu^+,\hat{X}_{\nu^\prime}^-]=2i\delta_{\nu \nu^\prime}$, and uncertainty relation $V(\hat{X}_\nu^+)V(\hat{X}_\nu^-)\geq 1$.

We firstly consider an amplitude squeezed atomic pulse propagating into the interaction region, with a weak optical probe field (linear intensity $1.9\times 10^{-7} m^{-1}$) incident on the condensate. This defines the transverse mode of the emitted probe photons. The atomic, pump, and probe wavevectors are $2k_0$, $-k_0$, $k_0$ respectively, with $k_{0}= 8\times 10^6 m^{-1}$, giving an atom laser beam velocity of $v_{atom} = 1.1\;cms^{-1}$. The input pulse $L_\psi(x,0)$ is a Gaussian of width $\sigma_x= 100\mu m$ containing $n_0=5\times 10^3$ atoms. We use $N_0=10^6$ condensate atoms, trapped with frequency $\omega_t=5$ Hz. In all cases we operate at the optimal efficiency point for the signal so that the ratio of the condensate width to the mean beam velocity is tuned to one quarter of a Rabi cycle, $T_{\rm Rabi}\approx 4\sqrt{\hbar/m\omega_t}(m/2\hbar k_0)$~\cite{Haine2006}.
The input atom laser pulse is modelled as an $X_{\psi}^{+}$-squeezed minimum uncertainty state with $V(\hat{X}_{\psi}^\pm)=e^{\mp 2r}$. The field intensities are shown in Fig.~\ref{fig:packetTimeSeries} (a) when the pulse is almost half incoupled. When the probe is initially in the vacuum state there is no outcoupling and the incoupling is almost perfectly efficient. In the presence of a continuous weak probe (shown here) some outcoupling also occurs. The probe quadrature variances for the same system are plotted in Fig.~\ref{fig:packetTimeSeries} (b). The incident probe light and the outcoupling dynamics have a negligible effect on the quadrature signal when compared with the vacuum probe case (identical on this scale). The squeezing of the input pulse is transmitted to the probe, allowing the quadrature statistics of the atom laser to be read out via optical homodyne detection of the probe light. 
%================================================
\begin{figure}[t]\begin{centering}\includegraphics[width=1\columnwidth]{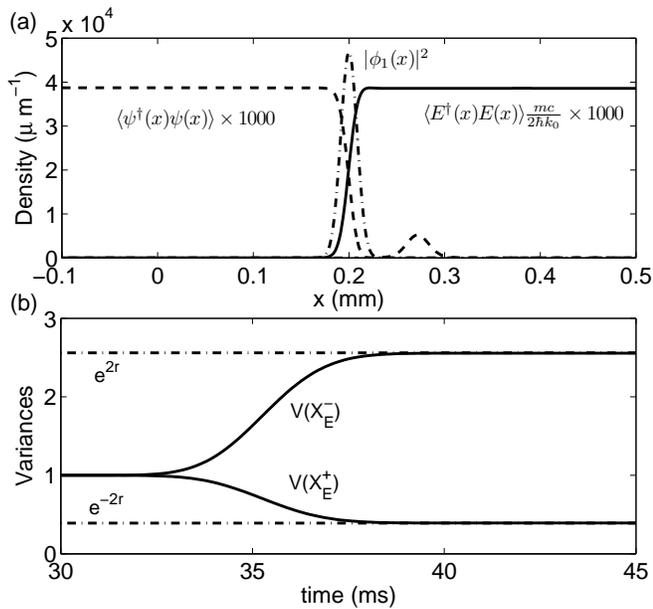}\par\end{centering}

\vspace*{-10pt}

\caption{\label{fig:BeamTimeSeries}\textbf{Incoupling a continuous atom laser beam}. (a) A snapshot of the atomic beam (dashed line), condensate (chain line) and probe (solid line) during incoupling. (b) The development of the optical quadratures (solid lines) as the front of the atom laser beam crosses the interaction region. The zero of the time axis is arbitrary. The chain lines show the variances of the $4$dB squeezed atomic beam.}

\vspace*{-10pt}

\end{figure}
%===============================================
This results demonstrates a limitation of pulsed dynamics: the squeezed quadrature of the probe shows less squeezing than the input atomic pulse. While the effect is not always large, it is significant for the chosen scenario because the spatial extent of the squeezed pulse is much larger than the condensate. The signal degrades because different parts of the same atomic wavepacket are subject to different Rabi frequencies whilst travelling across the interaction region.  
\par
For continuous, essentially monochromatic, squeezed atom laser input~\cite{Haine2005} the spatial effects are removed and the scheme efficiently maps the physical variances of the beam to the probe. Fig.~\ref{fig:BeamTimeSeries} shows the Raman incoupling dynamics. The system now consists of a squeezed atom laser beam in a nearly monochromatic state (wavevector $2k_0$) which enters from the left. The atoms are incoupled via the reversed Raman scheme, emitting probe photons. Once the front of the beam crosses the interaction region the system is approximately in a steady state (except for the gradual transfer of atoms into the trapped condensate), with a constant probe output. We see that the quadrature variances of the emitted probe light reach steady state values very close to the atom laser variances. 
\par
We now show analytically that the local $g^{(2)}$ can be extracted from the probe field with high efficiency. In fact we derive a more general result relating $g_\nu^{(2)}(x,x,t)\equiv\langle \hat{\nu}^\dag(x,t)\hat{\nu}^\dag(x,t)\hat{\nu}(x,t)\hat{\nu}(x,t)\rangle/\langle\hat{\nu}^\dag(x,t)\hat{\nu}(x,t)\rangle^2$ to the initial state of the atom laser beam and the probe field. Since the system is linear we may introduce a linear ansatz for the field operators $\hat{\nu}(x,t)=f_\nu(x,t) \hat{a}_0 + h_\nu(x,t) \hat{b}_0$ where the evolution of $f_\nu, h_\nu$ gives the field time development and the initial states are given by the single-mode bosonic operators $\hat{a}_0$ (atoms) and $\hat{b}_0$~\cite{method}. For the systems we consider one of $\langle \hat{b}_0\rangle$ or $\langle \hat{a}_0\rangle$ is zero, and the atom laser beam and optical probe are initially uncorrelated. We immediately find
\begin{widetext}
\vspace*{-15pt}
\begin{equation}
g_\nu^{(2)}(x,x,t)=\frac{|f_\nu(x,t)|^4\langle \hat{a}_0^\dag\hat{a}_0^\dag\hat{a}_0\hat{a}_0\rangle+|h_\nu(x,t)|^4\langle \hat{b}_0^\dag\hat{b}_0^\dag\hat{b}_0\hat{b}_0\rangle + 4|f_\nu(x,t)|^2|h_\nu(x,t)|^2\langle \hat{a}_0^\dag\hat{a}_0\rangle\langle \hat{b}_0^\dag\hat{b}_0\rangle}{|f_\nu(x,t)|^4\langle \hat{a}_0^\dag\hat{a}_0\rangle^2+|h_\nu(x,t)|^4\langle \hat{b}_0^\dag\hat{b}_0\rangle^2 + 2|f_\nu(x,t)|^2|h_\nu(x,t)|^2\langle \hat{a}_0^\dag\hat{a}_0\rangle\langle \hat{b}_0^\dag\hat{b}_0\rangle}
\end{equation}
\end{widetext}
where the field density $\langle \hat{\nu}(x,t)\hat{\nu}(x,t)\rangle$ is non-zero. The case of special interest here is when the optical probe is initially in the vacuum state so that $h_\nu(x,t)\equiv h_\nu(x,0)=0$ and $g_\nu^{(2)}(x,x,t)=\langle \hat{a}_0^\dag\hat{a}_0^\dag\hat{a}_0\hat{a}_0\rangle/\langle \hat{a}_0^\dag\hat{a}_0\rangle^2$, where defined. This demonstrates that $g_\psi^{(2)}(x,x,t)$ is mapped to the emitted probe light. Since this correlation can be directly measured using existing techniques~\cite{Ottl2005} this provides an experimental consistency check.
\par
Apart from the stability of the lasers used, the main sources of possible signal degradation are spontaneous emission losses and phase noise due to atomic collisions. We will firstly look at spontaneous emission from the excited atomic level. 
The loss rate can be estimated from the spontaneous emission rate for a transition with energy $\omega_0=k_0 c$ radiating into a continuum, $\gamma_{\rm sp}=k_0^3|d_{13}|^2/3\pi\hbar\epsilon_0$.
The total spontaneous loss during the incoupling is then
$L_{\rm sp}=\gamma_{\rm sp}\int dx\;\int dt\; \langle \hat{\psi}_3^\dag(x,t)\hat{\psi}_3(x,t)\rangle$.
Using the adiabatically eliminated expression for the excited state $\langle \hat{\psi}_3^\dag(x,t)\hat{\psi}_3(x,t)\rangle\approx\langle \hat{\psi}_2^\dag(x,t)\hat{\psi}_2(x,t)\rangle(\Omega_{23}/\Delta)^{2}$, and the fact that each excited atom on average remains excited for time $T_{\rm Rabi}/4$, we have $L_{\rm sp}\lesssim\gamma_{\rm sp}\bar{N}_3T_{\rm Rabi}/4$, where $\bar{N}_3$ is the total number of excited state atoms transferred per squeezed mode.
For the incoupling process to remain coherent, we require
$L_{\rm sp}/N_2\ll 1$, and upon integration over the entire input pulse for our parameters we find $L_{\rm sp}/N_2\approx 0.04$. We can now estimate the effect on the signal phenomenologically using a beam splitter which mixes the signal and vacuum with reflectivity $\eta$ ($\approx 0.04$ here). The probe variances then become $V(\hat{X}_E^\pm)=(1-\eta)V(\hat{X}_\psi^\pm)+\eta$, acceptable for small $\eta$. 

The effect of atomic collisions will be greatest within the trapped condensate, but as we are transferring the statistics of the input field to the probe light, these will have little effect. The collisions between the incoming and the trapped atoms will have two undesired effects. Firstly, there will be a mean-field effect which will tend to rotate the quadrature phases. This can be compensated for during the homodyne measurement stage. The second effect will be that of phase-diffusion of the beam, which to a first approximation will cause an increase in the variance of the phase quadrature. As the interaction time over which this can happen may be kept short, this should also not be a fatal drawback. Another issue which will arise is that the probe beam will be emitted into a narrow cone rather than as a well collimated beam. This can be simply overcome using linear optical elements. Finally, treating the BEC as a coherent state assumes that the atom laser beam and the BEC are phase correlated. One way this can be achieved is by dividing a larger BEC into two; the beam extracted from the first BEC will then be phase correlated with the second BEC used for Raman incoupling.

We have shown that a Raman incoupler scheme may be used as a means to measure the quantum statistics of an atom laser by transferring quadrature variances to an optical probe on which standard homodyne measurements may be made. Experimental realisation of our proposal would allow access to new quantum features of matter fields, including demonstrations of squeezing, entanglement between atomic beams, atom-light entanglement, the EPR paradox with matter waves.
%=========================================================================
\begin{acknowledgments}
We thank Matthew Davis, Peter Drummond and Yvan Castin for stimulating input.
This work was supported by the Australian Research Council.
\end{acknowledgments}
%=========================================================================

\end{document}